\documentclass[a4paper]{article}

\newcommand\blfootnote[1]{%
  \begingroup
  \renewcommand\thefootnote{}\footnote{#1}%
  \addtocounter{footnote}{-1}%
  \endgroup
}

\usepackage{INTERSPEECH2019}

\title{Interpretable Deep Learning Model for the Detection and Reconstruction of Dysarthric Speech}

\name{Daniel Korzekwa$^1$, Roberto Barra-Chicote$^1$, Bozena Kostek$^2$, Thomas Drugman$^1$, Mateusz Lajszczak$^1$}
\address{
$^1$Amazon TTS-Research\\
$^2$ Gdansk University of Technology, Faculty of ETI, Poland}
\email{korzekwa@amazon.com, rchicote@amazon.com, bokostek@multimed.org, drugman@amazon.com, mateuszl@amazon.com}

\begin{document}

\maketitle
\begin{abstract}
\blfootnote{Paper accepted on Interspeech 2019}We present a novel deep learning model for the detection and reconstruction of dysarthric speech. We train the model with a multi-task learning technique to jointly solve dysarthria detection and speech reconstruction tasks. The model key feature is a low-dimensional latent space that is meant to encode the properties of dysarthric speech. It is commonly believed that neural networks are “black boxes” that solve problems but do not provide interpretable outputs. On the contrary, we show that this latent space successfully encodes interpretable characteristics of dysarthria, is effective at detecting dysarthria, and that manipulation of the latent space allows the model to reconstruct healthy speech from dysarthric speech. This work can help patients and speech pathologists to improve their understanding of the condition, lead to more accurate diagnoses and aid in reconstructing healthy speech for afflicted patients.
\end{abstract}
\noindent\textbf{Index Terms}: dysarthria detection, speech recognition, speech synthesis, interpretable deep learning models

\section{Introduction}

Dysarthria is a motor speech disorder manifesting itself by a weakness
of muscles controlled by the brain and nervous system that are used in the process of speech production, such as lips, jaw and throat \cite{ASHA2018}. Patients with dysarthria produce harsh and breathy speech with abnormal prosodic
patterns, such as very low speech rate or flat intonation, which makes
their speech unnatural and difficult to comprehend. Damage to the nervous system is the main cause of dysarthria \cite{ASHA2018}.
It can happen as an effect of multiple possible neurological disorders
such as cerebral palsy, brain stroke, dementia or brain cyst \cite{Cuny2017,Banovic2018}.

Early onset detection of dysarthria may improve the quality of life
for people affected by these neurological disorders. According to Alzheimer's Research UK2015 \cite{Alzheimersresearchuk2015}, 1 out of 3 people
in the UK born in 2015 will develop dementia in their life. 
Manual detection of dysarthria conducted 
in clinical conditions by speech pathologists is costly, time-consuming and can lead to an incorrect diagnosis \cite{2012E121001,Carmichael2008}. With an automated analysis of speech, we can detect an early onset of
dysarthria and recommend further health checks with a clinician even when a human speech pathologist is not available. Speech reconstruction may help with better identification of the symptoms and
enable patients with severe dysarthria to communicate with other people.

Section 2 presents related work. In Section 3 we describe the proposed model for detection
and reconstruction of dysarthria. In Section 4 we demonstrate the
performance of the model with experiments on detection, interpretability, and
reconstruction of healthy speech from dysarthric speech. We conclude with our remarks.

\section{Related work}

\subsection{Dysarthria detection}

Deep neural networks can automatically detect dysarthric patterns
without any prior expert knowledge \cite{Krishna,DBLP:conf/interspeech/Vasquez-CorreaA18}.
Unfortunately, these models are difficult to interpret because
they are usually composed of multiple layers producing multidimensional outputs
with an arbitrary meaning and representation. Contrarily, statistical
models based on a fixed vector of handcrafted prosodic and spectral
features such as jitter, shimmer, Noise to Harmonic Ratio (NHR) or
Mel-Frequency Cepstral Coefficients (MFCC) offer good interpretability but require experts to manually design predictor features \cite{Falk2012,Sarria-Paja2012a,Gillespie2017,Lansford2014}.

The work of Tu Ming et al. on interpretable objective evaluation of dysarthria
\cite{DBLP:conf/interspeech/TuBL17} is the closest we found to our
proposal. The main difference is that our model not only provides interpretable characteristics of dysarthria but also reconstructs healthy speech. Their model is based
on feed-forward deep neural networks with a latent layer representing
four dimensions of dysarthria: nasality, vocal quality, articulatory
precision, and prosody. The final output of the network represents
general dysarthria severity on a scale from 1 to 7. The input to this
model is described by a 1201-dimensional vector of spectral
and cepstral features that capture various aspects of dysarthric
speech such as rhythm, glottal movement or formants. As opposed to
this work, we use only mel-spectrograms to present the input speech
to the model. Similarly to our approach, Vasquez-Correa et al. \cite{DBLP:conf/interspeech/Vasquez-CorreaA18}
uses a mel-spectrogram representation for dysarthria detection. However,
they use 160 ms long time windows at the transition points between
voiced and unvoiced speech segments, in contrast to using a full mel-spectrogram
in our approach.

\subsection{Speech reconstruction}

There are three different approaches to the reconstruction of dysarthric speech: voice banking, voice adaptation and voice reconstruction \cite{2012E121001}. Voice banking is a simple idea of collecting  a patient's speech samples before their speech becomes unintelligible and using it to build a personalized Text-To-Speech (TTS) voice.
It requires about 1800 utterances for a basic unit-selection TTS
technology \cite{Modeltalker} and more than 5K utterances for building a Neural TTS voice \cite{DBLP:journals/corr/abs-1811-06315}. Voice adaptation requires as little as
7 minutes of recordings. In this approach, we start with a TTS
model of an average speaker and adapt its acoustic and articulatory
parameters to the target speaker \cite{AhmadKhan2011}. 

Both voice banking and voice adaptation techniques rely on the availability of recordings for a healthy speaker. The voice reconstruction technique overcomes this shortcoming.
This technique aims at restoring damaged speech by tuning parameters representing the glottal source and the vocal tract filter \cite{rabiner_schafer78,Drugman2014}.
In our model, we
take a similar approach. However, instead of making assumptions on
what parameters should be restored, we let the model automatically learn
the best dimensions of the latent space that are responsible for dysarthric
speech. Reconstruction of healthy speech by manipulating the latent space of a dysarthric speech is a promising direction,
however, so far we only managed to successfully apply this technique in a single-speaker setup. 

Variational Auto-Encoder (VAE) \cite{doersch2016tutorial}
is a probabilistic latent space model that
has recently become popular for the reconstruction of various signals
such as text \cite{DBLP:journals/corr/HuYLSX17,DBLP:journals/corr/BowmanVVDJB15}
and speech \cite{DBLP:journals/corr/abs-1812-04342,DBLP:journals/corr/abs-1709-07902}.

\section{Proposed model}

\begin{figure}[ht]
  \centering
  \includegraphics[width=\linewidth]{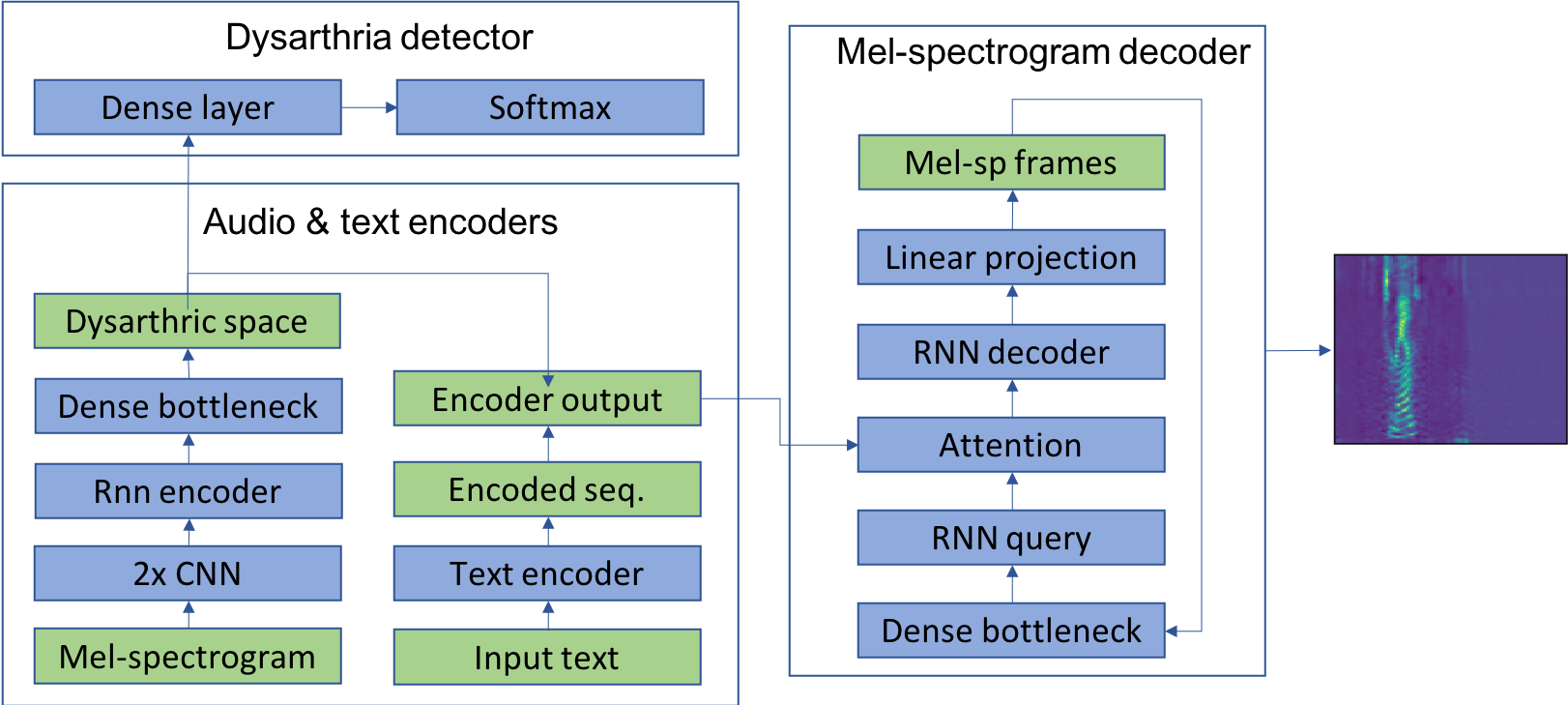}
  \caption{Architecture of deep learning model for detection and reconstruction
of dysarthric speech.}
  \label{fig:Architecute-of-deep}
\end{figure}

The model consists of two output networks, jointly trained, with a
shared encoder as shown in Figure \ref{fig:Architecute-of-deep}. The audio and text encoders produce a low-dimensional dysarthric latent space and a sequential encoding of the input text.
The audio decoder reconstructs input mel-spectrogram from a dysarthric latent space and encoded text. Logistic classification model predicts the probability of dysarthric speech from the dysarthric latent space. In Table \ref{tab:Configuration-of-neural}
we present the details of various neural blocks used in the model.

Let us define a matrix \textbf{$X:[n_{mels},n_{f}]$} representing a mel-spectrogram (frame length=50ms and frame shift=12.5ms), where
 \emph{$n_{mels}=128$} is the number of mel-frequency
bands and \emph{$n_{f}$} is the number of frames. Let us define a matrix $T:[n_{c},n_{t}]$ representing a one-hot encoded input text, where $n_{c}$ is the number of unique characters in the
alphabet and $n_{t}$ is the number of characters in the input text.
The mel-spectrogram $X$ is encoded into 2-dimensional
dysarthria latent space $\mathbf{l}=\{l_{1},l_{2}\}$ and then used as a conditioning variable for estimating the probability 
of dysarthria $d \backsim p(d|X,\theta)$ and reconstructing the mel-spectrogram $Y \backsim p(Y|X,T,\theta)$. Limiting the latent space to 2 dimensions makes the model more resilient to overfitting. The theta is a vector of trainable parameters of the model.

Let us define a training set of $m$ tuples of $((X,T),y)$, where $y\in\{0,1\}$ is the label for normal/dysarthric
speech and $m$ is the number of speech mel-spectrograms for dysarthric
and normal speakers. We optimize a joint cost of the predicted probability
of dysarthria and mel-spectrogram reconstruction defined as a weighted
function:
\begin{equation}
\sum_{i=1}^{m}\alpha log(p(d_{i}|X_{i},\theta))+(1-\alpha)log(p(Y_{i}|X_{i},T_{i},\theta))\label{eq:}
\end{equation}
where $log(p(d_{i}|X_{i},\theta))$ is the cross-entropy between the predicted
and actual labels of dysarthria, and $log(p(Y_{i}|X_{i},T_{i},\theta))$ is the log-likelihood
of a Gaussian distribution for the predicted mel-spectrogram with a unit
variance, a.k.a L2 loss. We used backpropagation and mini-batch stochastic gradient
descent with a learning rate of 0.03 and a batch size of 50.
The whole model is initialized with Xavier’s method \cite{journals/jmlr/GlorotB10} using the magnitude
value of 2.24. Hyper-parameters of the model presented in Table \ref{tab:Configuration-of-neural} were tuned with a  grid search optimization. We used MxNet framework for implementing the model \cite{DBLP:journals/corr/ChenLLLWWXXZZ15}.

\begin{table}[t]
  \caption{Configuration of the neural network blocks.}
  \label{tab:Configuration-of-neural}
  \centering
  \begin{tabular}{ll}
    \toprule  
    \textbf{Neural block}  & \textbf{Config}                \\
    \midrule
    \multicolumn{2}{c}{\textbf{Audio encoder}} \\
    2x CNN & 20 channels, 5x5 kernel, RELU, VALID \\
    GRU & 20 hidden states, 1 layer \\
    Dense & 20 units, tanh \\
    Dysarthric space & 2 units, linear \\
    \multicolumn{2}{c}{\textbf{Text encoder}} \\
    3x CNN & 40 channels, 5x5 kernel, RELU, SAME \\
    GRU & 27 hidden states, 1 layer \\
    \multicolumn{2}{c}{\textbf{Audio decoder}} \\
    Dense bottleneck & 96 units, RELU \\
    GRU query & 29 hidden states, 1 layer \\
    GRU decoder & 128 hidden states, 1 layer \\
    Linear projection & frames\_num x melsp bins units, linear \\
   
    \bottomrule
  \end{tabular}

\end{table}

\subsection{Mel-spectrogram and text encoders}

For the spectrogram encoder, we use a Recurrent Convolutional Neural Network
model (RCNN) \cite{DBLP:journals/corr/abs-1803-09047}. The convolutional layers,
each followed by a max-pooling layer, extract local and time-invariant patterns of
the glottal source and the vocal tract. The GRU layer
models temporal patterns of dysarthric speech \cite{DBLP:journals/corr/ChoMGBSB14}. The last
state of the GRU layer is processed by two dense layers. Dropout \cite{JMLR:v15:srivastava14a} with probability of 0.5 is applied to the output of the activations for both
CNN layers, GRU layer, and the dense layer.

Text encoder encodes the input text using one-hot encoding, followed by
three CNN layers and one GRU layer. Outputs of both audio and text encoders are concatenated via matrix broadcasting, producing a matrix $E:[n_{c}+n_{l},n_{t}]$, where $n_{l}$ is dimensionality 
of the dysarthria latent space.

\subsection{Spectrogram decoder and dysarthria detector }

For decoding a mel-spectrogram, similarly to Wang et al. \cite{DBLP:journals/corr/WangSSWWJYXCBLA17}, we use a Recurrent Neural Network (RNN) model with attention. The
dot-product attention mechanism \cite{DBLP:journals/corr/VaswaniSPUJGKP17} plays a crucial role.
It informs to which elements of the encoder output the decoder should pay attention at every decoder step. The RNN network that produces a query vector for the attention, takes as input $r$ predicted mel-spectrogram frames from the previous time-step. 
The output of the RNN decoder is projected via a linear dense
layer into $r$ number of mel-spectrogram frames. Similarly to Wang
et al. \cite{DBLP:journals/corr/WangSSWWJYXCBLA17}, we found that
it is important to preprocess the mel-spectrogram with a dense layer and
dropout regularization to improve the overall generalization of the model.

The dysarthria detector is created from a 2-dimensional dense layer. It
uses a tanh activation followed by a softmax function that represents
the probability of dysarthric speech.

\section{Experiments}

\subsection{Dysarthric speech database\label{subsec:Dysarthric-speech-database}}

There is no well-established benchmark in the literature to compare
different models for detecting dysarthria. Aside from the most popular
dysarthric corpora, UA-Speech \cite{Kim2008} and TORGO \cite{Rudzicz2012}, there are multiple speech databases created for the purpose of a
specific study, for example, corpora of 57 dysarthric speakers \cite{Lansford2014}
and Enderby Frenchay Assessment dataset \cite{Carmichael2008}.
Many corpora, including TORGO and HomeService \cite{Nicolao2016}, are
available under non-commercial license.

In our experiments we use the UA-Speech database from the University
of Illinois \cite{Kim2008}. It contains 11 male and 4 female dysarthric
speakers of different dysarthria severity levels and 13 control speakers.
455 isolated words are recorded for each speaker with 1 to 3 repetitions.
Every word is recorded through a 7-channel microphone array, producing a separate wav file of 16 kHz sampling rate
for every channel. It contains 9.4 hours of speech for dysarthric speakers
and 4.85 hours for control speakers. UA-Speech corpus comes with intelligibility scores that are obtained from a transcription task performed by 5 naive listeners.

To control variabilities in recording conditions, we normalized mel-spectrograms for every recorded
word independently with a z-score normalization. We considered removing the initial period of silence at the beginning
of recorded words but we decided against it. We found that for dysarthric
speakers of high speech intelligibility, the average length of the initial
silence period that lasts 0.569sec +- 0.04674 (99\% CI) is comparable with
healthy speakers with the length of 0.532sec +- 0.055. Because we can predict unvoiced periods with merely 85\% of accuracy \cite{Johnston:2012:WAR:2432294}, removing the periods of silence for dysarthric speakers with poor intelligibility is very inaccurate.

\subsection{Automatic detection of dysarthria\label{subsec:Automatic-detection-of}}

To define the training and test sets, we use a Leave-One-Subject-Out (LOSO) cross-validation scheme. For each
training, we include all speakers but one that is left out to measure
the prediction accuracy on unseen examples. The accuracy, precision and recall metrics are computed at a speaker level (the average dysarthria probability
of all the words produced by the speaker is compared to a target speaker dysarthria
label $\in\{0,1\}$), and a word level (comparing target dysarthria
label with predicted dysarthria probability for all words independently).

As a baseline, we use the Gillespie's et al. model that is based on Support Vector Machine classifier \cite{Gillespie2017}.
It uses 1595 low-level predictor features processed with a global z-score normalization.
It reports a 75.3 and 92.9 accuracy in the dysarthria detection
task at the word and speaker levels respectively, following LOSO cross-validation. However, Gillespie uses 336
words from the UA-Speech corpus with 12 words per speaker, whereas we
use all 455 words across all speakers. 

In our first model, only dysarthric labels are observed and we achieved an accuracy on the word and speaker
levels of 82\% and 93\% respectively. By training the multi-task model,
in which both targets, i.e. mel-spectrogram and dysarthric labels, are observed, the accuracy on the word level increased by 3 percents to the value of 85.3\% (Table \ref{tab:Accuracy-metrics-for}). We found that the UA-Speech database includes multiple recorded words for healthy speakers that contain intelligibility errors, different words than asked or background speech of other people. These
issues affect the accuracy of detecting dysarthric speech.

\begin{table}[ht]
  \caption{Accuracy of dysarthria detection including  95\% CI. Classifier task - target mel-spectrogram (ML) is not
observed during training. Multitask - both targets ML and dysarthric
labels are observed}
  \label{tab:Accuracy-metrics-for}
  \centering
  \begin{tabular}{llll}
    \toprule  
    \textbf{System} & \textbf{Accuracy} & \textbf{Precision}  & \textbf{Recall}\\
    \midrule
    \multicolumn{4}{c}{\textbf{Word level}} \\
    Multitask & 0.853 (0.849 - 0.857) & 0.831 & 0.911\\
    Classifier task & 0.820 (0.815 - 0.824) & 0.818 & 0.855 \\
    Gillespie et al.\cite{Gillespie2017} & 0.753 (na) & 0.823 & 0.728 \\
    \multicolumn{4}{c}{\textbf{Speaker level}} \\
    Multitask & 0.929 (0.790-0.984) & 1.000 & 0.867\\
    Classifier task & 0.929 (0.790-0.984) & 0.933 & 0.933 \\
    Gillespie et al.\cite{Gillespie2017} & 0.929 (na) & na & na \\
    \bottomrule
  \end{tabular}
\end{table}

Krishna reports a 97.5\% accuracy on UA-Corpus \cite{Krishna}. However,
after email clarification with the author, we found that they estimated the accuracy
taking into account only the speakers with a medium level of dysarthria.
Narendra et al. achieved 93.06\% utterance level accuracy on the
TORGO dysarthric speech database \cite{DBLP:conf/interspeech/NarendraA18}. As opposed to the related work, our model does not need any expert knowledge to design hand-crafted features and it can learn automatically using a low-dimensional latent space that encodes characteristics of dysarthria.

\subsection{Interpretable modeling of dysarthric patterns\label{subsec:Interpretable-modeling-of}}

We analyze the correlation between the dysarthric
latent space and the intelligibility of speakers. We look at 550 audio samples
of a single 'Command' word across the 15 dysarthric
speakers and 13 healthy speakers.

\begin{figure}[t]
  \centering
  \includegraphics[width=\linewidth,height=5cm]{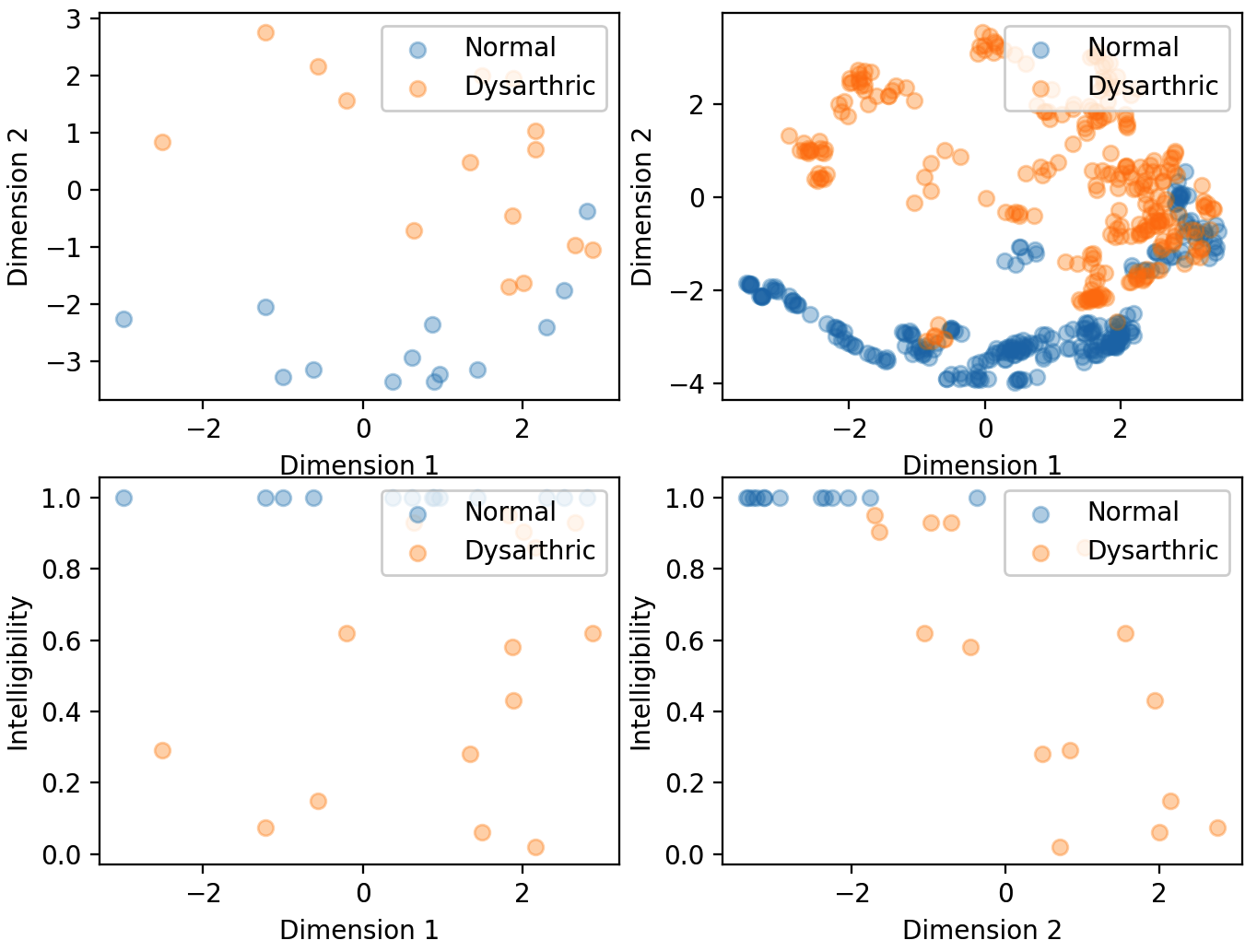}
  \caption{Unsupervised learning. Top row: Separation between dysarthric and control speakers in the latent space on a speaker (left) and word (right) level. Bottom row: Correlation between both dimensions of the latent space and the intelligibility scores.}
  \label{fig:Correlation-between-dysarthric}
\end{figure}

\begin{figure}[t]
  \centering
  \includegraphics[width=\linewidth,height=5cm]{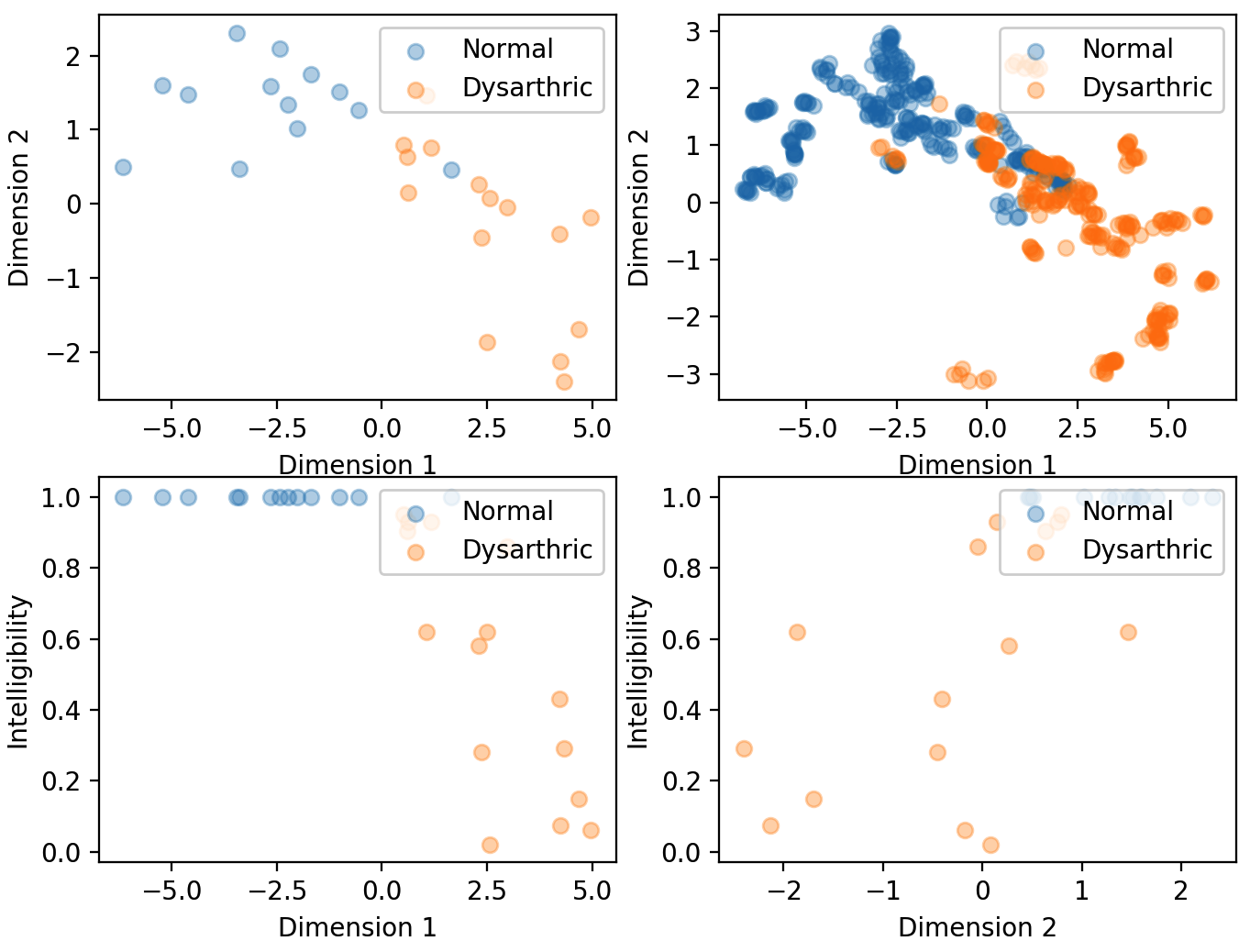}
  \caption{Supervised learning. As in Figure \ref{fig:Correlation-between-dysarthric}.}
  \label{fig:Correlation-between-dysarthric-1}
\end{figure}

In an unsupervised training (Figure \ref{fig:Correlation-between-dysarthric}), target labels of dysarthric/normal speech are not presented
to the model. Dysarthric speakers are well separated from normal speakers and the
dimension 2 of the latent space is negatively correlated with the
intelligibility scores (Pearson correlation of -0.84, two-sided p-value \verb|<| 0.001).
In a supervised variant (Figure \ref{fig:Correlation-between-dysarthric-1}), we train the model jointly
with both reconstructed mel-spectrogram and the target dysarthria labels
observed. Both dimensions of the latent space are highly correlated with the
intelligibility scores (dimension 1 with correlation of -0.76 and
dimension 2 with correlation of 0.70, both with p-value \verb|<| 0.001). 

The sign of the correlation has no particular meaning. Retraining the model multiple times results in both positive and negative correlations between the latent space and the intelligibility of speech.
A high correlation 
between dysarthric latent space and intelligibility
scores suggests that by moving along the dimensions of the latent space,
we should be able to reconstruct speech of dysarthric speakers and
improve its intelligibility. We explore this in the next experiment.

\subsection{Reconstruction of dysarthric speech\label{subsec:Reconstruction-of-dysarthric}}

First we trained a supervised multi-speaker model with all dysarthric and control speakers but
we achieved poor reconstruction results with almost unintelligible speech. We think
this is due to a high variability of dysarthric speech across all speakers,
including various articulation, prosody and fluency problems. To better understand the potential for speech reconstruction, we narrowed the experiment down to two speakers, male speaker M05 and a corresponding control speaker. We have chosen M05 subject because their speech varies across different levels of fluency and we wanted to observe
this pattern when manipulating the latent space. For example, when pronouncing the word 'backspace', M05 uttered consonants 'b' and 's' multiple times, resulting in 'ba ba cs space'.

We analyzed a single category of 19 computer command words,
such as 'command' or 'backspace'. For every word uttered by M05,
we generated 5 different versions of speech, fixing dimension 2 of
the latent space to the value of -0.1, and using the values of {[}-0.5,
0, 0.5, 1, 1.5{]} for dimension 1. Audio samples of reconstructed
speech were obtained by converting predicted mel-spectrograms to waveforms
using the Griffin-Lim algorithm \cite{Griffin1984}.

We conducted MUSHRA perceptual test \cite{Merritt2018}.
Every listener was presented with 6 versions of a
given word at the same time, 5 reconstructions and one version of
recorded speech. We asked listeners to evaluate the fluency of speech
on a scale from 0 to 100. We used
10 US based listeners from the Amazon mTurk platform,
in total providing us with 1140 evaluated speech samples.

\begin{figure}[t]
  \centering
  \includegraphics[width=\linewidth,height=3cm]{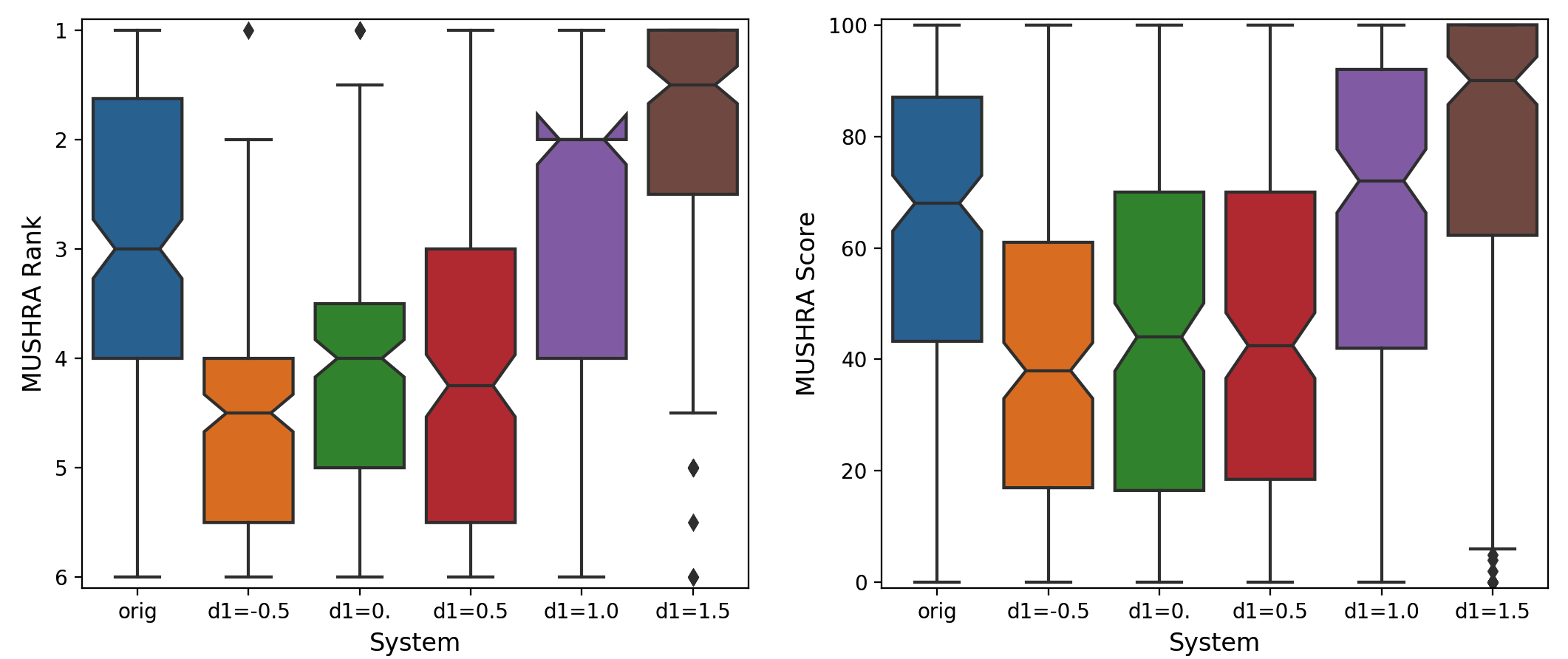}
  \caption{MUSHRA results for the fluency of speech for 5 reconstructions and one recorded speech. Rank order (left) and
the median score on the scale from 0 to 100 (right).}
  \label{fig:MUSHRA-results-for}
\end{figure}

\begin{figure}[t]
  \centering
  \includegraphics[width=\linewidth]{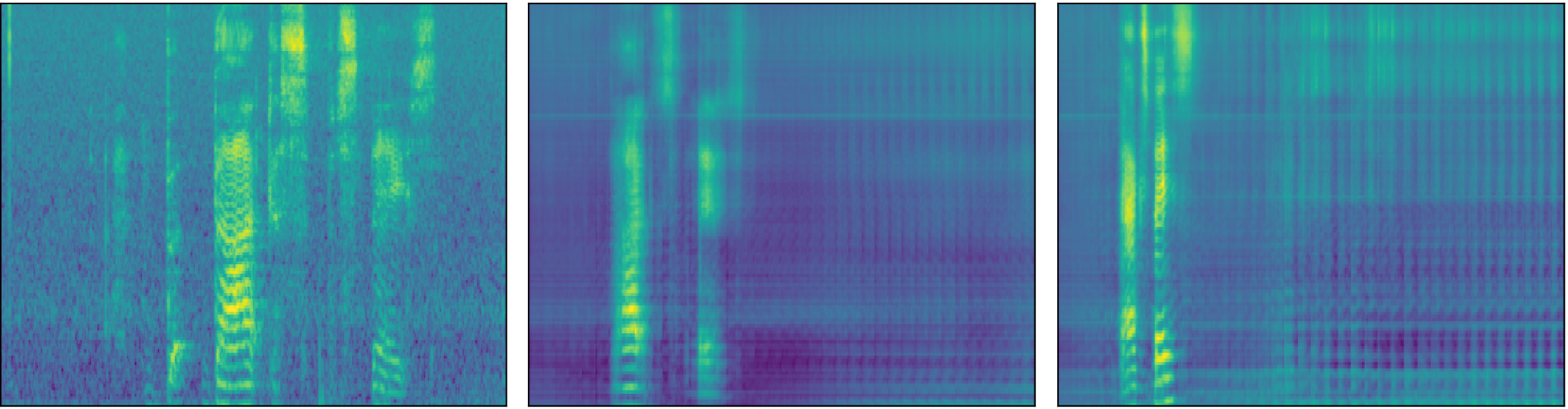}
  \caption{Reconstruction of dysarthric speech ('command' word). From left to right (MUSHRA scores of 51.8, 61.9 and 89.5): Recorded dysarthric speech. Reconstructed speech
with dimension 1 of 0.0 and 1.5 respectively.}
  \label{fig:Reconstruction-of-dysarthric}
\end{figure}

As shown in Figure \ref{fig:MUSHRA-results-for}, by moving along dimension 1 of the latent space, we
can improve the fluency of speech, generating speech with levels of fluency not observed in the training data.
In the pairwise two-sided Wilcoxon signed-rank, all pairs of ranks are different
from each other with p-value \verb|<| 0.001, except of \{orig, d1=1.0\}, \{d1=-0.5,
d1=0.0\}, \{d1=-0.5, d1=0.5\}. Examples of original and reconstructed mel-spectrograms are shown in Figure \ref{fig:Reconstruction-of-dysarthric}. 

We found that manipulation of the latent space changes both the fluency of speech and the timbre of voice and it is possible that dysarthria is so tied up with speaker identify making it fruitless to disentangle them. We replaced a deterministic dysarthric latent space with a Gaussian variable and trained the model with an additional Kullback\verb|-|Leibler loss \cite{doersch2016tutorial,Mathieu2018} but we did not manage to separate the timbre of voice from dysarthria. Training the model with an additional discriminative cost to ensure that every dimension of the latent space is directly associated with a particular speech factor can potentially help with this problem \cite{DBLP:journals/corr/HuYLSX17}. 

\section{Conclusions}
 
This paper proposed a novel approach for the detection and reconstruction of dysarthric speech. The encoder-decoder model factorizes speech into a low-dimensional latent space and encoding of the input text. We showed that the latent space conveys interpretable characteristics of dysarthria, such as intelligibility and fluency of speech. MUSHRA perceptual test demonstrated that the adaptation of the latent space let the model generate speech of improved fluency. The multi-task supervised approach for predicting both the probability of dysarthric speech and the mel-spectrogram helps improve the detection of dysarthria with higher accuracy. This is thanks to a low-dimensional latent space of the auto-encoder as opposed to directly predicting dysarthria from a highly dimensional mel-spectrogram.
 
\section{Acknowledgements}

We would like to thank A. Nadolski, J. Droppo, J. Rohnke and V. Klimkov for insightful discussions on this work. 

\bibliographystyle{IEEEtran}

\bibliography{mybib}

\begin{thebibliography}{10}
\providecommand{\url}[1]{#1}
\csname url@samestyle\endcsname
\providecommand{\newblock}{\relax}
\providecommand{\bibinfo}[2]{#2}
\providecommand{\BIBentrySTDinterwordspacing}{\spaceskip=0pt\relax}
\providecommand{\BIBentryALTinterwordstretchfactor}{4}
\providecommand{\BIBentryALTinterwordspacing}{\spaceskip=\fontdimen2\font plus
\BIBentryALTinterwordstretchfactor\fontdimen3\font minus
  \fontdimen4\font\relax}
\providecommand{\BIBforeignlanguage}[2]{{%
\expandafter\ifx\csname l@#1\endcsname\relax
\typeout{** WARNING: IEEEtran.bst: No hyphenation pattern has been}%
\typeout{** loaded for the language `#1'. Using the pattern for}%
\typeout{** the default language instead.}%
\else
\language=\csname l@#1\endcsname
\fi
#2}}
\providecommand{\BIBdecl}{\relax}
\BIBdecl

\bibitem{ASHA2018}
ASHA, ``{The American Speech-Language-Hearing Association (ASHA) -
  Dysarthria},'' 2018.

\bibitem{Cuny2017}
M.~L. Cuny, M.~Pallone, H.~Piana, N.~Boddaert, C.~Sainte-Rose,
  L.~Vaivre-Douret, P.~Piolino, and S.~Puget, ``{Neuropsychological improvement
  after posterior fossa arachnoid cyst drainage},'' \emph{Child's Nervous
  System}, 2017.

\bibitem{Banovic2018}
\BIBentryALTinterwordspacing
S.~Banovic, L.~Zunic, and O.~Sinanovic, ``{Communication Difficulties as a
  Result of Dementia},'' \emph{Materia Socio Medica}, vol.~30, no.~2, p. 221,
  2018. [Online]. Available:
  \url{https://www.ejmanager.com/fulltextpdf.php?mno=302643414}
\BIBentrySTDinterwordspacing

\bibitem{Alzheimersresearchuk2015}
Alzheimersresearchuk, ``{One in three people born in 2015 will develop
  dementia, new analysis shows},'' 2015.

\bibitem{2012E121001}
J.~Yamagishi, C.~Veaux, S.~King, and S.~Renals, ``{Speech synthesis
  technologies for individuals with vocal disabilities: Voice banking and
  reconstruction},'' \emph{Acoustical Science and Technology}, vol.~33, no.~1,
  pp. 1--5, 2012.

\bibitem{Carmichael2008}
J.~Carmichael, V.~Wan, and P.~Green, ``{Combining neural network and rule-based
  systems for dysarthria diagnosis},'' in \emph{Proceedings of the Annual
  Conference of the International Speech Communication Association,
  INTERSPEECH}, 2008.

\bibitem{Krishna}
G.~Krishna, ``{Excitation Source Analysis of Dysarthric Speech for Early Stage
  Detection of Dysarthria},'' \emph{WSPD}, 2018.

\bibitem{DBLP:conf/interspeech/Vasquez-CorreaA18}
J.~C. V{\'{a}}squez-Correa, T.~Arias-Vergara, J.~R. Orozco-Arroyave, and
  E.~N{\"{o}}th, ``{A Multitask Learning Approach to Assess the Dysarthria
  Severity in Patients with Parkinson's Disease},'' in \emph{Interspeech 2018,
  19th Annual Conference of the International Speech Communication Association,
  Hyderabad, India, 2-6 September 2018.}, B.~Yegnanarayana, Ed.\hskip 1em plus
  0.5em minus 0.4em\relax ISCA, 2018, pp. 456--460.

\bibitem{Falk2012}
T.~H. Falk, W.~Y. Chan, and F.~Shein, ``{Characterization of atypical vocal
  source excitation, temporal dynamics and prosody for objective measurement of
  dysarthric word intelligibility},'' \emph{Speech Communication}, 2012.

\bibitem{Sarria-Paja2012a}
M.~Sarria-Paja and T.~Falk, ``{Automated Dysarthria Severity Classification for
  Improved Objective Intelligibility Assessment of Spastic Dysarthric
  Speech.}'' in \emph{Interspeech}, 2012.

\bibitem{Gillespie2017}
S.~Gillespie, Y.~Y. Logan, E.~Moore, J.~Laures-Gore, S.~Russell, and R.~Patel,
  ``{Cross-database models for the classification of dysarthria presence},'' in
  \emph{Proceedings of the Annual Conference of the International Speech
  Communication Association, INTERSPEECH}, 2017.

\bibitem{Lansford2014}
K.~L. Lansford and J.~M. Liss, ``{Vowel Acoustics in Dysarthria: Speech
  Disorder Diagnosis and Classification},'' \emph{Journal of Speech Language
  and Hearing Research}, 2014.

\bibitem{DBLP:conf/interspeech/TuBL17}
M.~Tu, V.~Berisha, and J.~Liss, ``{Interpretable Objective Assessment of
  Dysarthric Speech Based on Deep Neural Networks},'' in \emph{Interspeech
  2017, 18th Annual Conference of the International Speech Communication
  Association, Stockholm, Sweden, August 20-24, 2017}, F.~Lacerda, Ed.\hskip
  1em plus 0.5em minus 0.4em\relax ISCA, 2017, pp. 1849--1853.

\bibitem{Modeltalker}
Modeltalker, ``www.modeltalker.com.''

\bibitem{DBLP:journals/corr/abs-1811-06315}
J.~Latorre, J.~Lachowicz, J.~Lorenzo-Trueba, T.~Merritt, T.~Drugman,
  S.~Ronanki, and K.~Viacheslav, ``{Effect of data reduction on
  sequence-to-sequence neural {\{}TTS{\}}},'' \emph{CoRR}, vol. abs/1811.0,
  2018.

\bibitem{AhmadKhan2011}
Z.~{Ahmad Khan}, P.~Green, S.~Creer, and S.~Cunningham, ``{Reconstructing the
  voice of an individual following laryngectomy},'' 2011.

\bibitem{rabiner_schafer78}
L.~Rabiner and R.~Schafer, \emph{{Digital Processing of Speech Signals}}.\hskip
  1em plus 0.5em minus 0.4em\relax Englewood Cliffs: Prentice Hall, 1978.

\bibitem{Drugman2014}
T.~Drugman, P.~Alku, A.~Alwan, and B.~Yegnanarayana, ``Glottal source
  processing: from analysis to applications,'' \emph{Computer Speech and
  Language}, vol.~28, 09 2014.

\bibitem{doersch2016tutorial}
C.~Doersch, ``{Tutorial on Variational Autoencoders},'' 2016.

\bibitem{DBLP:journals/corr/HuYLSX17}
Z.~Hu, Z.~Yang, X.~Liang, R.~Salakhutdinov, and E.~P. Xing, ``{Controllable
  Text Generation},'' \emph{CoRR}, vol. abs/1703.0, 2017.

\bibitem{DBLP:journals/corr/BowmanVVDJB15}
S.~R. Bowman, L.~Vilnis, O.~Vinyals, A.~M. Dai, R.~J{\'{o}}zefowicz, and
  S.~Bengio, ``{Generating Sentences from a Continuous Space},'' \emph{CoRR},
  vol. abs/1511.0, 2015.

\bibitem{DBLP:journals/corr/abs-1812-04342}
Y.-J. Zhang, S.~Pan, L.~He, and Z.-H. Ling, ``{Learning latent representations
  for style control and transfer in end-to-end speech synthesis},''
  \emph{CoRR}, vol. abs/1812.0, 2018.

\bibitem{DBLP:journals/corr/abs-1709-07902}
W.-N. Hsu, Y.~Zhang, and J.~R. Glass, ``{Unsupervised Learning of Disentangled
  and Interpretable Representations from Sequential Data},'' \emph{CoRR}, vol.
  abs/1709.0, 2017.

\bibitem{journals/jmlr/GlorotB10}
X.~Glorot and Y.~Bengio, ``Understanding the difficulty of training deep
  feedforward neural networks.'' in \emph{AISTATS}, ser. JMLR Proceedings,
  Y.~W. Teh and D.~M. Titterington, Eds., vol.~9.\hskip 1em plus 0.5em minus
  0.4em\relax JMLR.org, 2010, pp. 249--256.

\bibitem{DBLP:journals/corr/ChenLLLWWXXZZ15}
\BIBentryALTinterwordspacing
T.~Chen, M.~Li, Y.~Li, M.~Lin, N.~Wang, M.~Wang, T.~Xiao, B.~Xu, C.~Zhang, and
  Z.~Zhang, ``Mxnet: {A} flexible and efficient machine learning library for
  heterogeneous distributed systems,'' \emph{CoRR}, vol. abs/1512.01274, 2015.
  [Online]. Available: \url{http://arxiv.org/abs/1512.01274}
\BIBentrySTDinterwordspacing

\bibitem{DBLP:journals/corr/abs-1803-09047}
R.~J. Skerry-Ryan, E.~Battenberg, Y.~Xiao, Y.~Wang, D.~Stanton, J.~Shor, R.~J.
  Weiss, R.~Clark, and R.~A. Saurous, ``{Towards End-to-End Prosody Transfer
  for Expressive Speech Synthesis with Tacotron},'' \emph{CoRR}, vol.
  abs/1803.0, 2018.

\bibitem{DBLP:journals/corr/ChoMGBSB14}
K.~Cho, B.~van Merrienboer, {\c{C}}.~G{\"{u}}l{\c{c}}ehre, F.~Bougares,
  H.~Schwenk, and Y.~Bengio, ``{Learning Phrase Representations using
  {\{}RNN{\}} Encoder-Decoder for Statistical Machine Translation},''
  \emph{CoRR}, vol. abs/1406.1, 2014.

\bibitem{JMLR:v15:srivastava14a}
N.~Srivastava, G.~Hinton, A.~Krizhevsky, I.~Sutskever, and R.~Salakhutdinov,
  ``{Dropout: A Simple Way to Prevent Neural Networks from Overfitting},''
  \emph{Journal of Machine Learning Research}, vol.~15, pp. 1929--1958, 2014.

\bibitem{DBLP:journals/corr/WangSSWWJYXCBLA17}
Y.~Wang, R.~J. Skerry-Ryan, D.~Stanton, Y.~Wu, R.~J. Weiss, N.~Jaitly, Z.~Yang,
  Y.~Xiao, Z.~Chen, S.~Bengio, Q.~V. Le, Y.~Agiomyrgiannakis, R.~Clark, and
  R.~A. Saurous, ``{Tacotron: {\{}A{\}} Fully End-to-End Text-To-Speech
  Synthesis Model},'' \emph{CoRR}, vol. abs/1703.1, 2017.

\bibitem{DBLP:journals/corr/VaswaniSPUJGKP17}
A.~Vaswani, N.~Shazeer, N.~Parmar, J.~Uszkoreit, L.~Jones, A.~N. Gomez,
  L.~Kaiser, and I.~Polosukhin, ``{Attention Is All You Need},'' \emph{CoRR},
  vol. abs/1706.0, 2017.

\bibitem{Kim2008}
H.~Kim, M.~Hasegawa-Johnson, A.~Perlman, J.~Gunderson, T.~Huang, K.~Watkin, and
  S.~Frame, ``{Dysarthric Speech Database for Universal Access Research},''
  \emph{INTERSPEECH}, 2008.

\bibitem{Rudzicz2012}
F.~Rudzicz, A.~K. Namasivayam, and T.~Wolff, ``{The TORGO database of acoustic
  and articulatory speech from speakers with dysarthria},'' \emph{Language
  Resources and Evaluation}, 2012.

\bibitem{Nicolao2016}
M.~Nicolao, H.~Christensen, S.~Cunningham, P.~Green, and T.~Hain, ``{A
  framework for collecting realistic recordings of dysarthric speech - The
  homeService corpus},'' in \emph{Proceedings of the 10th International
  Conference on Language Resources and Evaluation, LREC 2016}, 2016.

\bibitem{Johnston:2012:WAR:2432294}
A.~B. Johnston and D.~C. Burnett, \emph{WebRTC: APIs and RTCWEB Protocols of
  the HTML5 Real-Time Web}.\hskip 1em plus 0.5em minus 0.4em\relax USA: Digital
  Codex LLC, 2012.

\bibitem{DBLP:conf/interspeech/NarendraA18}
N.~P. Narendra and P.~Alku, ``{Dysarthric Speech Classification Using Glottal
  Features Computed from Non-words, Words and Sentences},'' in
  \emph{Interspeech 2018, 19th Annual Conference of the International Speech
  Communication Association, Hyderabad, India, 2-6 September 2018.},
  B.~Yegnanarayana, Ed.\hskip 1em plus 0.5em minus 0.4em\relax ISCA, 2018, pp.
  3403--3407.

\bibitem{Griffin1984}
D.~W. Griffin and J.~S. Lim, ``{Signal Estimation from Modified Short-Time
  Fourier Transform},'' \emph{IEEE Transactions on Acoustics, Speech, and
  Signal Processing}, 1984.

\bibitem{Merritt2018}
T.~Merritt, B.~Putrycz, A.~Nadolski, T.~Ye, D.~Korzekwa, W.~Dolecki,
  T.~Drugman, V.~Klimkov, A.~Moinet, A.~Breen, R.~Kuklinski, N.~Strom, and
  R.~Barra-Chicote, ``{Comprehensive evaluation of statistical speech waveform
  synthesis},'' nov 2018.

\bibitem{Mathieu2018}
E.~Mathieu, T.~Rainforth, N.~Siddharth, and Y.~W. Teh, ``{Disentangling
  Disentanglement in Variational Auto-Encoders},'' dec 2018.

\end{thebibliography}

\end{document}